\documentclass[10pt,letterpaper]{article}
\usepackage{opex3}
\usepackage{citesort}

\begin{document}

\title{Coherent Phonon Dynamics in Short-Period InAs/GaSb Superlattices}

\author{G.~T.~Noe,$^1$ H.~J.~Haugan,$^2$ G.~J.~Brown,$^2$ \\G.~D.~Sanders,$^3$ C.~J.~Stanton,$^3$ and J.~Kono$^{1,*}$}
\address{$^1$Department of Electrical and Computer Engineering, Rice University, \\ Houston, Texas 77005, USA
\\ $^2$Air Force Research Laboratory, Materials and Manufacturing Directorate, \\Wright-Patterson Air Force Base, Ohio 45433-7707, USA
\\$^3$Department of Physics, University of Florida, Gainesville, Florida 32611-8440, USA}

\email{kono@rice.edu} 



\begin{abstract} We have performed ultrafast pump-probe spectroscopy studies on a series of InAs/GaSb-based short-period superlattice (SL) samples with periods ranging from 46~\AA~to 71~\AA.  We observe two types of oscillations in the differential  reflectivity with fast ($\sim$ 1- 2 ps)  and slow ($\sim$ 24 ps) periods.  The period of the fast oscillations changes with the SL period and can be explained as coherent acoustic phonons generated from carriers photoexcited within the SL.  This mode provides an accurate method for determining the SL period and assessing interface quality. The period of the slow mode depends on the wavelength of the probe pulse and can be understood as a propagating coherent phonon wavepacket modulating the reflectivity of the probe pulse as it travels from the surface into the sample.
\end{abstract}

\ocis{(300.6495) Spectroscopy, terahertz; (320.7150) Ultrafast spectroscopy.} 



\section{Introduction}
\label{intro}
Development of quantum engineered materials and devices for high-performance sensing covering a wide spectral range in the infrared (IR) is important for a broad range of potential applications.  In particular, materials and structures are being sought that can be exploited to produce uncooled, or high-operating temperature, detectors for diagnostic and surveillance purposes.  Today's IR imaging arrays are based on either HgCdTe or intersubband quantum well photon detectors, but both types of detectors require cryogenic cooling in order to produce high sensitivities.
Short-period superlattices (SLs) based on alternating layers of InAs and GaSb are expected to be a promising material for IR detectors~\cite{Sai-halaszetAl77APL,SmithMailhiot87JAP,HauganetAl04APL}. Due to the type-II band alignment of these SLs, the band gap can be tuned to the mid-IR range by an appropriate choice of SL period.  Also, it is expected that the band structure of these SLs can be tailored to reduce Auger recombination and tunneling currents to make room temperature operation possible~\cite{GreinetAl92APL}.  However, growth and fabrication of high-quality detectors using these SLs still remains challenging, and basic materials properties, especially dynamical properties, are not well understood.

Here, we have performed ultrafast pump-probe spectroscopy on InAs/GaSb SLs to increase our understanding of dynamical properties such as  scattering and recombination rates.  We find that the differential reflectivity decays with a very fast ($\sim$1~ps) and a very slow ($>$~10~ns) time constant.  This is associated with intraband and interband dynamics of photo-excited carriers.  In addition, we observed two different types of oscillating behavior in the differential reflectivity spectrum which we associate with coherent phonons generated by the photoexcited carriers.  The first type --- slow oscillations with a period of $\sim$24~ps --- was commonly observed in all the SL samples and did not change with sample, whereas the second type --- fast oscillations with a period of $\sim$2~ps --- changed in frequency from sample to sample and depended on the SL period.  The fast oscillations can be explained as coherent acoustic phonons generated in the SL whose wave vector is determined by the SL period.  This mode provides an accurate method for determining the SL period and assessing interface quality~\cite{Marris98SciAm}.

\section{Samples}
\label{samples}
The five samples used in this study were short-period InAs/GaSb type-II SLs with varied SL periods: 46.0, 49.5, 56.3, 61.0, and 71.2~\AA.  All of the samples were grown by molecular beam epitaxy on Te-doped ($n$-type) GaSb (100) substrates.  Initial growth conditions were roughly tuned to achieve morphologically smooth surfaces using a 0.5~$\mu$m thick 21~\AA~InAs/24~\AA~GaSb SLs (45~\AA-SL), and the same growth conditions were applied throughout the wider SL samples.  In a typical growth procedure, a 0.3-$\mu$m-thick GaSb buffer layer was grown at 490$^{\circ}$C, followed by a 0.5-$\mu$m thick SL layer grown at 400$^{\circ}$C.  The V/III beam equivalent flux ratio was set around 3 for both InAs and GaSb layers, with a growth rate of 0.6~\AA/s for both InAs and GaSb layers.  Following the SL layer growth, the samples were annealed in-situ at 450$\pm$10$^{\circ}$C for 30 minutes under Sb-overpressure.  In order to insert controlled 1.6~\AA~InSb-like interfaces between the layers, a migration-enhanced epitaxy technique was used, and their sequential details can be found in~\cite{HauganetAl11JVST}.  The SL structural parameters were determined through high-resolution X-ray diffraction, and their details are listed in Table I.
\begin{table}
\begin{center}
\begin{tabular}{c|cc|c|c}
\hline\hline
Sample & InAs/GaSb (\AA/\AA) & SL Period (\AA) & SL Period (\AA) & SL Period (\AA) \\
No. & Nominal & Nominal & X-Ray & Coherent Phonons\\
\hline
1 & 21.0/24.0 & 45 & 46.0 & 46.0 \\
2 & 26.5/21.5 & 48 & 49.5 & 50.0 \\
3 & 33.5/21.5 & 55 & 56.3 & 58.2 \\
4 & 38.5/21.5 & 60 & 61.0 & 61.8 \\
5 & 48.5/21.5 & 70 & 71.2 & 68.6 \\
\hline\hline
\end{tabular}
\caption{\label{table}Data summary for the InAs/GaSb superlattice samples studied in this work.  Each layer includes a 1.6~\AA-thick InSb-like interface, and their nominal thicknesses were estimated from shutter times.  The third column shows the superlattice periods obtained from high resolution X-Ray diffraction, and the last column shows the superlattice periods obtained from coherent phonon measurements described in this paper.}
\end{center}
\end{table}

\section{Experimental Methods}\label{methods}
To investigate the carrier and phonon dynamics for these samples, we used standard pump-probe techniques and measured the differential reflectivity spectrum. To measure the dynamics of photo-generated carriers and phonons in a wide temporal range, we used a chirped pulse amplifier (CPA-2010, Clark-MXR, Inc.) with a wavelength of 795~nm, a repetition rate of 1~kHz, and a pulse-width of $\sim$150~fs.  This CPA seeds an optical parametric amplifier (OPA).  In this experiment, the OPA, tuned to 1.5~$\mu$m, was used as the pump, and the CPA was used as the probe.  The probe passed an optical delay line that consisted of two long linear stages for an overall delay of $\sim$8.3~ns.  The reflected probe was collected into a fast photodiode, and the differential signal was measured by a lock-in amplifier referenced to the optical chopper frequency which modulated the pump.

In order to measure the fast coherent phonon oscillations with small amplitudes (fractional reflectivity change $\Delta R/R$ $\sim$1~$\times$~10$^{-5}$), we used a higher repetition rate, $\sim$93~MHz, Ti:Sapphire oscillator laser.  We performed degenerate pump-probe measurements using the oscillator laser with a wavelength of 795~nm and a pulse-width of $\sim$50~fs.  The pump line passed an optical delay line that consisted of a short linear stage with a fast shaker.  A beam splitter was placed in the probe path, which provided a reference for a balanced detector to compare the reflected probe with the reference to eliminate any common mode noise.  All measurements were done at room temperature.

\section{Experimental Results}
\begin{figure}[h]
\centering
\includegraphics[scale=0.55]{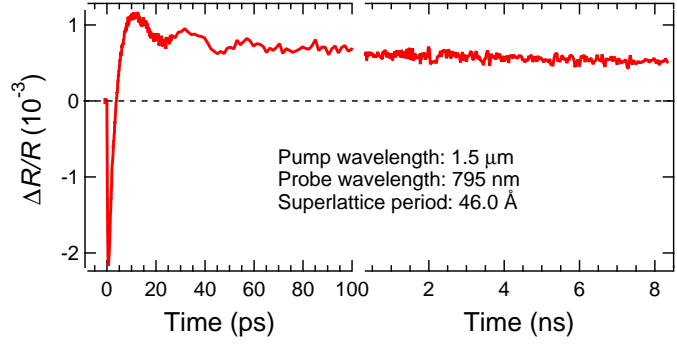}
\centering \caption{Overall pump-probe differential reflectivity dynamics for Sample 1 with a superlattice period of 46.0~\AA.  The normalized differential reflectivity shows three features: a sharp decrease followed by a sign change taking a few ps, slow oscillations with period of $\sim$24~ps, and a recovery much slower than 8~ns.}
\label{overall}
\end{figure}

We measured the overall time-dependent reflectivity using the CPA/OPA system for all five samples.  Figure~\ref{overall} shows the results in two time ranges for Sample 1 (with SL period of 46\AA).  Initially, the reflectivity shows a sharp decrease followed by a fast increase and a sign change on the order of a picosecond.  Then, there are slow oscillations with period of $\sim$24~ps. All five samples exhibit slow oscillations with the same oscillation period.  The slow oscillation results from a coherent phonon wavepacket being generated near the surface of the sample and modulating the reflectivity of the probe pulse as it propagates into the sample.  The theory for the slow oscillations has been worked out in Ref.~\cite{LiutAl05PRB}. Finally, there is a slow recovery that was not fully measured with an 8~ns optical delay.  This recovery to zero is too slow to be measured with standard delay stage techniques because a delay line of tens of meters would be required. However, there is no signal before zero delay, which means that there must be a recovery to zero before 1~ms (the spacing between pulses for a 1~kHz repetition rate laser).

\begin{figure}[h]
\centering\includegraphics[scale=0.5]{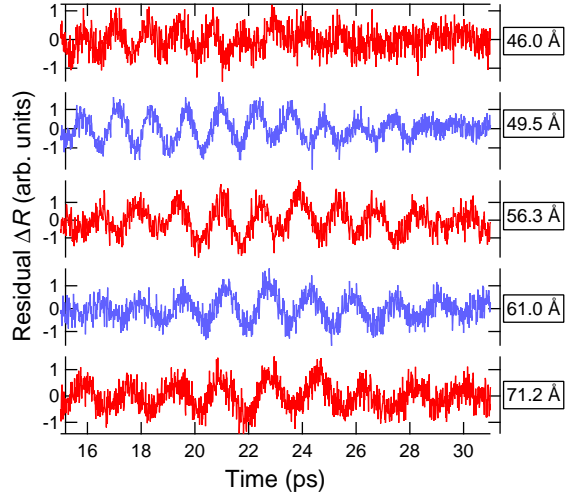}
\centering \caption{The residual differential reflectivity for different superlattice-period samples after subtracting the background signal to isolate the fast coherent phonon oscillations.  The oscillations show an increase in period with increasing SL period (shown on the right side).}
\label{fast}
\end{figure}

Figure~\ref{fast} shows the fast oscillations in the time-dependent reflectivity for five different samples using the 93~MHz Ti:Sapphire oscillator degenerate pump-probe setup with a pump/probe wavelength of 795~nm.  The raw data from this setup shows a time-dependent reflectivity signal that is similar to what was seen with the amplifier laser but with much better signal-to-noise ratio, allowing us to more easily resolve the faster oscillations.  The component of the differential reflectivity due to the photo-excited carries and the slow oscillation was subtracted by fitting the differential reflectivity with the sum of two exponential decays and two damped cosine functions.  The residual after the fit and subtraction shows only the \textit{fast} oscillations in the differential reflectivity.  We note that the period of the fast oscillations, unlike the slow oscillations, depends on the sample and increases systematically with SL period from $\sim$1~ps to $\sim$2~ps.

\begin{figure}[h]
\centering
\includegraphics[scale=0.45]{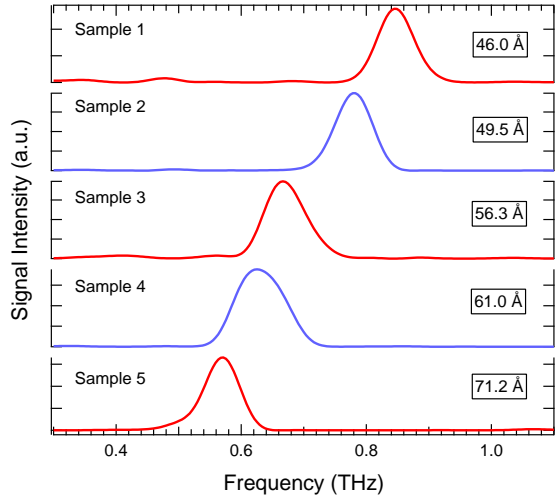}
\centering \caption{Fourier transform of the fast time-domain oscillations in Fig.~\ref{fast}.  The peak position of the oscillation frequency decreases with increasing superlattice period.}
\label{FFT}
\end{figure}

Figure~\ref{FFT} shows the results of a fast Fourier transform (FFT) of the residual time-domain data for all five samples.  Before taking the FFT, we used a linear interpolation to make 10,000 evenly spaced points.  We also used zero padding of 90,000 points to increase the data density of the FFT result.  We used a Hann window function.  Here, we see a large peak for each sample that corresponds to the frequency of the fast oscillations in reflectivity.  The frequency decreases with increasing SL period thickness from $\sim$0.84~THz to $\sim$0.58~THz.

\begin{figure}[h]
\centering
\includegraphics[scale=0.5]{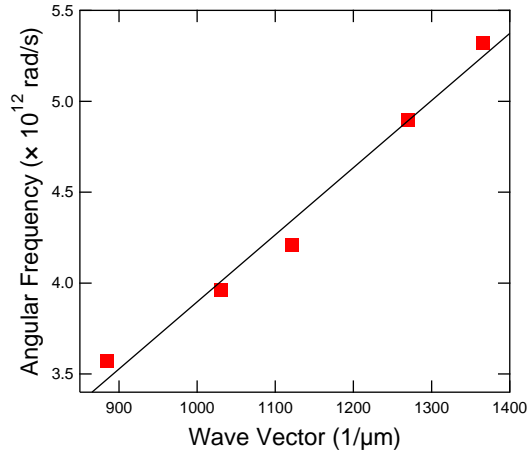}
\centering \caption{Angular frequency versus superlattice wave vector $q$.  The slope of the least squares fit corresponds to the sound velocity of 3.87~$\times$~10$^{5}$~cm/s.}
\label{fig4}
\end{figure}

\section{Discussion}

Because InAs/GaSb superlattices are type-II, photoexicted electrons and holes are confined in different layers.  As a result, the electron and hole wavefunctions have small overlap and InAs/GaSb-based short-period SLs are expected to demonstrate a very long carrier lifetime, on the order of tens of nanoseconds~\cite{DonetskyetAl09APL}.  The slowest feature shown in Fig.~\ref{overall} can be attributed to the recombination of optically excited carriers back across the gap to the original ground state.  Our results demonstrate that our samples have a lifetime much greater than the maximum delay time, 8~ns, that we could reasonably achieve using standard optical delay-line techniques.  Beam walk and beam divergence are two major problems associated with introducing an arbitrarily long optical delay.  Without measuring the full recovery to zero, we could not accurately determine the carrier lifetime using this method of measurement.  

The ultrafast generation of electrons and holes in the sample by the pump leads to generation of coherent longitudinal acoustic (LA) phonons near the surface that propogate into the sample~\cite{SandersetAl05PRB}.  The period of oscillation is given by $T= \lambda/[2C_{s}n(\lambda)]$~\cite{WangetAl05PRB}, where $\lambda$ is the wavelength of the probe, $C_{s}$ is the LA sound speed, and $n(\lambda)$ is the wavelength-dependent refractive index.  For Sample 1, we use the weighted average of the InAs and GaSb properties to calculate $C_{s}$ and $n$(795~nm).  Using these values, we calculate the period of oscillation to be 24.8~ps, which is close to the observed period of 24~ps. Note that the period of the slow oscillations does not depend on the SL period, but instead depends on the wavelength of the probe pulse.

The fast coherent phonon oscillations are more interesting since their frequencies are sample dependent. The fast oscillations arise from the non-uniform absorption in the InAs and GaSb layers.  The band gap at room temperature is substantially smaller in InAs (0.35 eV) than in GaSb (0.726 eV).  As a result, photoexcitation by either a 1.5 $\mu$m or 795 nm pump pulse will create substantially more electrons and holes in the InAs layers.  (The holes will subsequently rapidly transfer to the GaSb layers).  This leads to an electron or hole density matrix $n_q$ with a non-zero Fourier component $q$. These carriers can then couple to coherent phonons with frequencies $\omega(q)$ as shown in Ref.~\cite{kuznetovetal94PRL}.  The photo-excited carriers are coupled to the acoustic-phonon mode with wave vector $q=2\pi/L$, where $L$ is the SL period~\cite{Sanders01.235316,sunetal99APL}.  This effect is enhanced because of the type-II superlattice.

While some people might view these phonons as \textit{zone-folded} acoustic phonons, the (100) phonons in InAs and GaSb have nearly the same sound velocity (3.83~$\times~10^5$~cm/s for InAs vs.~3.97~$\times~10^5$~cm/s for GaSb), and, as a result, the reflection coefficient at the InAs/GaSb interface is small ($\approx 3 \times 10^{-4})$.  Hence, the phonons are almost bulk like.  \textit{It is the non-uniform photo-excitation of electrons and holes that triggers the coherent phonon oscillations.}

Figure 4 shows the angular frequency vs. superlattice wave vector $q$ for all five samples.  The angular frequency was determined by using a gaussian fit of the fourier transform to determine the center frequency.  The data fits well with a straight line and the slope corresponds to a sound velocity of 3.87~$\times$~10$^{5}$~cm/s.  Since the frequency of the fast mode depends on the sample, it can be used for assessing the superlattice period and the quality of the interfaces. 

\section{Conclusions}
We have performed time resolved differential reflectivity measurements on InAs/GaSb superlattices. In addition to fast ($\sim$~1~ps) and slow ($>$~8~ns) \textit{decay} times associated with the photoexcited carrier dynamics, we observe fast ($\sim$~2~ps)  and slow ($\sim$~24~ps) \textit{oscillations} associated with the generation of coherent phonons by the photoexicted carriers.  The slow oscillations are sample independent and result from the generation of coherent phonon wavepackets near the surface that modulate the reflectivity of the probe beam as they travel into the sample.  The frequency of the fast oscillations is sample dependent and related to the superlattice period $L$.  They arise from the different absorption coefficients in the InAs and GaSb layers.  This leads to a photoexcited carrier density that has a Fourier component at $q = 2\pi/L$ that can trigger a coherent phonon at that wavevector. Our results indicate that this fast oscillation provides a very accurate method 
for determining the SL period and assessing the quality of the superlattices.

\section{Acknowledgments}
This work was supported by the CONTACT Program and the National Science Foundation through grants OISE-0968405 and DMR-0706313.  We thank the Aspen Center for Physics where part of this work was performed.


\begin{thebibliography}{10}
\newcommand{\enquote}[1]{``#1''}

\bibitem{Sai-halaszetAl77APL}
G.~A. Sai-Halasz, R.~Tsu, and L.~Esaki, \enquote{A New Semiconductor
  Superlattice,} Appl. Phys. Lett. {\bf 30}, 651 (1977).

\bibitem{SmithMailhiot87JAP}
D.~L. Smith and C.~Mailhiot, \enquote{Proposal for Strained Type II
  Superlattice Infrared Detectors,} J. Appl. Phys. {\bf 62}, 2545 (1987).

\bibitem{HauganetAl04APL}
H.~J. Haugan, F.~Szmulowicz, G.~J. Brown, and K.~Mahalingam, \enquote{Band Gap
  Tuning of InAs/GaSb Type-II Superlattices for Mid-Infrared Detection,} J.
  Appl. Phys. {\bf 96}, 2580 (2004).

\bibitem{GreinetAl92APL}
C.~H. Grein, P.~M. Young, and H.~Ehrenreich, \enquote{Minority Carrier
  Lifetimes in Ideal InGaSb/InAs Superlattices,} Appl. Phys. Lett. {\bf 61},
  2905 (1992).

\bibitem{Marris98SciAm}
H.~Marris, \enquote{Picosecond Ultrasonics,} Scientific America, {\bf January},
  86 (1998).

\bibitem{HauganetAl11JVST}
H.~J. Haugan, G.~J. Brown, and L.~Grazulis, \enquote{Effect of Interfacial
  Formation on the Properties of Very Long Wavelength Infrared InAs/GaSb
  Superlattices,} J. Vac. Sci. Tech. B {\bf 29}, 03C101 (2011).

\bibitem{LiutAl05PRB}
R.~Liu, G.~D. Sanders, C.~J. Stanton, C.~S. Kim, J.~S. Yahng, Y.~D. Jho, K.~J.
  Yee, E.~Oh, and D.~S. Kim, \enquote{Femtosecond pump-probe spectroscopy of
  propagating coherent acoustic phonons in In$_x$Ga$_{1-x}$N/GaN
  heterostructures,} Phys. Rev. B {\bf 72}, 195335 (2005).

\bibitem{DonetskyetAl09APL}
D.~Donetsky, S.~P. Svensson, L.~E. Vorobjev, and G.~Belenky, \enquote{Carrier
  Lifetime Measurements in Short-Period InAs/GaSb Strained-Layer Superlattice
  Structures,} Appl. Phys. Lett. {\bf 95}, 212104 (2009).

\bibitem{SandersetAl05PRB}
G.~D. Sanders, C.~J. Stanton, J.~Wang, C.~Sun, J.~Kono, A.~Oiwa, and
  H.~Munekata, \enquote{Theory of Carrier Dynamics and Time-Resolved
  Reflectivity in InMnAs/GaSb Heterostructures,} Phys. Rev. B {\bf 72}, 245302
  (2005).

\bibitem{WangetAl05PRB}
J.~Wang, Y.~Hashimoto, J.~Kono, A.~Oiwa, H.~Munekata, G.~D. Sanders, and C.~J.
  Stanton, \enquote{Propagating Coherent Acoustic Phonon Wave Packets in
  InMnAs/GaSb,} Phys. Rev. B {\bf 72}, 153311 (2005).

\bibitem{kuznetovetal94PRL}
A.~V. Kuznetsov and C.~J. Stanton, \enquote{Theory of Coherent Phonon
  Oscillations in Semiconductors,} Phys. Rev. Lett. {\bf 73}, 3243 (1994).

\bibitem{Sanders01.235316}
G.~D. Sanders, C.~J. Stanton, and C.~S. Kim, \enquote{Theory of Coherent
  Acoustic Phonons in InGaN/GaN Multi-Quantum Wells,} Phys. Rev. B {\bf 64},
  235316 (2001).

\bibitem{sunetal99APL}
C.-K. Sun, J.-C. Liang, A.~Abare, L.~Coldren, S.~P. Denbaars, and C.~J.
  Stanton, \enquote{Large Coherent Acoustic-Phonon Oscillations in InGaN/GaN
  Multiple-Quantum Wells,} Appl. Phys. Lett. {\bf 75}, 1249 (1999).

\end{thebibliography}
\end{document}